\documentclass[journal=jacsat,manuscript=article]{achemso}

\usepackage[version=4]{mhchem} 
\usepackage{textcomp}
\usepackage{upgreek}

\author{Yixiong Ji}
\affiliation{ARC Centre of Excellence in Exciton Science, School of Chemistry, University of Melbourne, Victoria 3010, Australia}

\author{Wentong Yang}
\affiliation{School of Metallurgy and Environment, Central South University, Changsha 410083, China}

\author{Di Yan}
\affiliation{Department of Electrical and Electronic Engineering, University of Melbourne, Victoria 3010, Australia}

\author{Wei Luo}
\affiliation{ARC Centre of Excellence in Exciton Science, School of Chemistry, University of Melbourne, Victoria 3010, Australia}

\author{Jialu Li}
\affiliation{ARC Centre of Excellence in Exciton Science, School of Chemistry, University of Melbourne, Victoria 3010, Australia}

\author{Shi Tang}
\affiliation{ARC Centre of Excellence in Exciton Science, School of Chemistry, University of Melbourne, Victoria 3010, Australia}

\author{Jintao Fu}
\affiliation{Chongqing Institute of Green and Intelligent Technology, Chinese Academy of Sciences, Chongqing 400714, China}

\author{James Bullock}
\affiliation{Department of Electrical and Electronic Engineering, University of Melbourne, Victoria 3010, Australia}

\author{Mei Gao}
\affiliation{CSIRO Manufacturing Clayton, Victoria 3168, Australia}

\author{Xin Li}
\affiliation{Chongqing Institute of Green and Intelligent Technology, Chinese Academy of Sciences, Chongqing 400714, China}

\author{Zhancheng Li}
\affiliation{Chongqing Institute of Green and Intelligent Technology, Chinese Academy of Sciences, Chongqing 400714, China}

\author{Jun Yang}
\affiliation{Chongqing Institute of Green and Intelligent Technology, Chinese Academy of Sciences, Chongqing 400714, China}

\author{Xingzhan Wei}
\affiliation{Chongqing Institute of Green and Intelligent Technology, Chinese Academy of Sciences, Chongqing 400714, China}

\author{Haofei Shi}
\affiliation{Chongqing Institute of Green and Intelligent Technology, Chinese Academy of Sciences, Chongqing 400714, China}

\author{Fangyang Liu}
\email{liufangyang@csu.edu.cn}
\affiliation{School of Metallurgy and Environment, Central South University, Changsha 410083, China}

\author{Paul Mulvaney}
\email{mulvaney@unimelb.edu.au}
\affiliation{ARC Centre of Excellence in Exciton Science, School of Chemistry, University of Melbourne, Victoria 3010, Australia}

\title[An \textsf{achemso} demo]
{A quasi-ohmic back contact achieved by inserting single-crystal graphene in flexible Kesterite solar cells}
\abbreviations{Solar energy materials}
\begin{document}

\begin{tocentry}

Some journals require a graphical entry for the Table of Contents.
This should be laid out ``print ready'' so that the sizing of the
text is correct.

Inside the \texttt{tocentry} environment, the font used is Helvetica
8\,pt, as required by \emph{Journal of the American Chemical
Society}.

The surrounding frame is 9\,cm by 3.5\,cm, which is the maximum
permitted for  \emph{Journal of the American Chemical Society}
graphical table of content entries. The box will not resize if the
content is too big: instead it will overflow the edge of the box.

This box and the associated title will always be printed on a
separate page at the end of the document.

\end{tocentry}


\newpage
\begin{abstract}

Flexible photovoltaics with a lightweight and adaptable nature that allows for deployment on curved surfaces and in building facades have always been a goal vigorously pursued by researchers in thin-film solar cell technology. The recent strides made in improving the sunlight-to-electricity conversion efficiency of kesterite Cu$_{2}$ZnSn(S, Se)$_{4}$ (CZTSSe) suggest it to be a perfect candidate. However, making use of rare Mo foil in CZTSSe solar cells causes severe problems in thermal expansion matching, uneven grain growth, and severe problems at the back contact of the devices. Herein, a strategy utilizing single-crystal graphene to modify the back interface of flexible CZTSSe solar cells is proposed. It will be shown that the insertion of graphene at the Mo foil/CZTSSe interface provides strong physical support for the subsequent deposition of the CZTSSe absorber layer, improving the adhesion between the absorber layer and the Mo foil substrate. Additionally, the graphene passivates the rough sites on the surface of the Mo foil, enhancing the chemical homogeneity of the substrate, and resulting in a more crystalline and homogeneous CZTSSe absorber layer on the Mo foil substrate. The detrimental reaction between Mo and CZTSSe has also been eliminated. Through an analysis of the electrical properties, it is found that the introduction of graphene at the back interface promotes the formation of a quasi-ohmic contact at the back contact, decreasing the back contact barrier of the solar cell, and leading to efficient collection of charges at the back interface. This investigation demonstrates that solution-based CZTSSe photovoltaic devices could form the basis of cheap and flexible solar cells.

\end{abstract}

\newpage
\section{Introduction}
 
Concerns are growing with the significant impacts of carbon dioxide and other greenhouse gases on the climate. The quest for renewable energy has been intensely improved in recent decades to meet the carbon-neutral criteria \cite{5-3}. The emerging development of solar energy is becoming a leading candidate due to its abundance and sustainability. Among various solar technologies, Cu$_{2}$ZnSn(S, Se)$_{4}$ (CZTSSe) solar cells have garnered significant attention owing to their element sufficiency and potential for flexible applications. These characteristics make CZTSSe solar cells appealing for various applications, from building-integrated photovoltaics (BIPV) to portable outdoor PV panels \cite{5-2}.

Significant progress has been made in the past decade on flexible CZTSSe solar cells \cite{5-4,5-5}. However, the efficiency still lags behind that of other flexible candidates and is also lower than rigid CZTSSe solar cells fabricated on Mo-coated soda lime glass. The disparity in the performance between the flexible and the rigid CZTSSe devices might be attributed to the absence of alkali metal and growing absorbers without contamination from the substrate, \cite{5-6,5-8,5-21}. The first problem can be solved by strategies like directly doping alkali metals into the precursors or through post-treatment have been demonstrated useful in compensating the alkali metals \cite{5-10,5-9,5-7}. However, strategies are still needed to solve the second problem.

In addition, rear/back contact in flexible CZTSSe solar cells suffers not only the problems that rigid ones have, but also severe characteristics such as thermal expansion mismatch, higher surface roughness, and detrimental element diffusion from substrates, which are critical in influencing the grain growth of absorbers and carrier transporting properties at the bottom contacts \cite{5-13,5-11,5-12,5-15,5-16,5-17}. The rear interface in flexible kesterite frequently exhibits issues due to voids along the Mo/CZTSSe interface. These voids are sometimes caused by the diffusion of various elements during the selenization process and the formation of volatile compounds like Sn(S, Se)$_2$ \cite{5-18,5-19}. Additionally, there is broad agreement on the harmful decomposition reaction at the back interface, which leads to the breakdown of the CZTSSe layer into Cu$_2$(S, Se), Zn(S, Se), and Sn(S, Se)$_2$, along with the growth of Mo(S, Se)$_2$ \cite{5-20,5-21}. The strategy of inserting materials used in rigid CZTSSe solar cells can help to solve the problem at the back contact. However, the materials often fall short when applied to flexible counterparts, resulting in decreased performance due to less flexibility and distinct thermal expansion during the thermal process. Therefore, novel materials are needed to be placed on flexible substrates for CZTSSe.

Herein, to fully exploit the advantages of flexible CZTSSe technology, single-crystal graphene with excellent thermal stability, conductivity, and flexibility was introduced as an intermediate layer, aiming to optimize the grain growth and carrier transporting at the back contact. Through a detailed examination of experimental results and theoretical analyses, this paper aims to highlight the significant impact of single-crystal graphene on back contact modifications, efficiency, and longevity of flexible CZTSSe solar cells. The incorporation of graphene helps to adjust the stress distribution at the back interface during the selenization, preventing collapse and uneven crystal growth caused by excessive thermal expansion. Additionally, the existence of graphene provides a quasi-ohmic electrical contact through graphene/Mo(S, Se)$_{2}$/Mo heterojunction, which is beneficial for carrier collection. It offers insights into future directions and potential research avenues that could further advance flexible CZTSSe devices and promote its applications.

\newpage
\section{Results and discussion}

\begin{figure}[H]
    \centering
    \includegraphics[width=\textwidth]{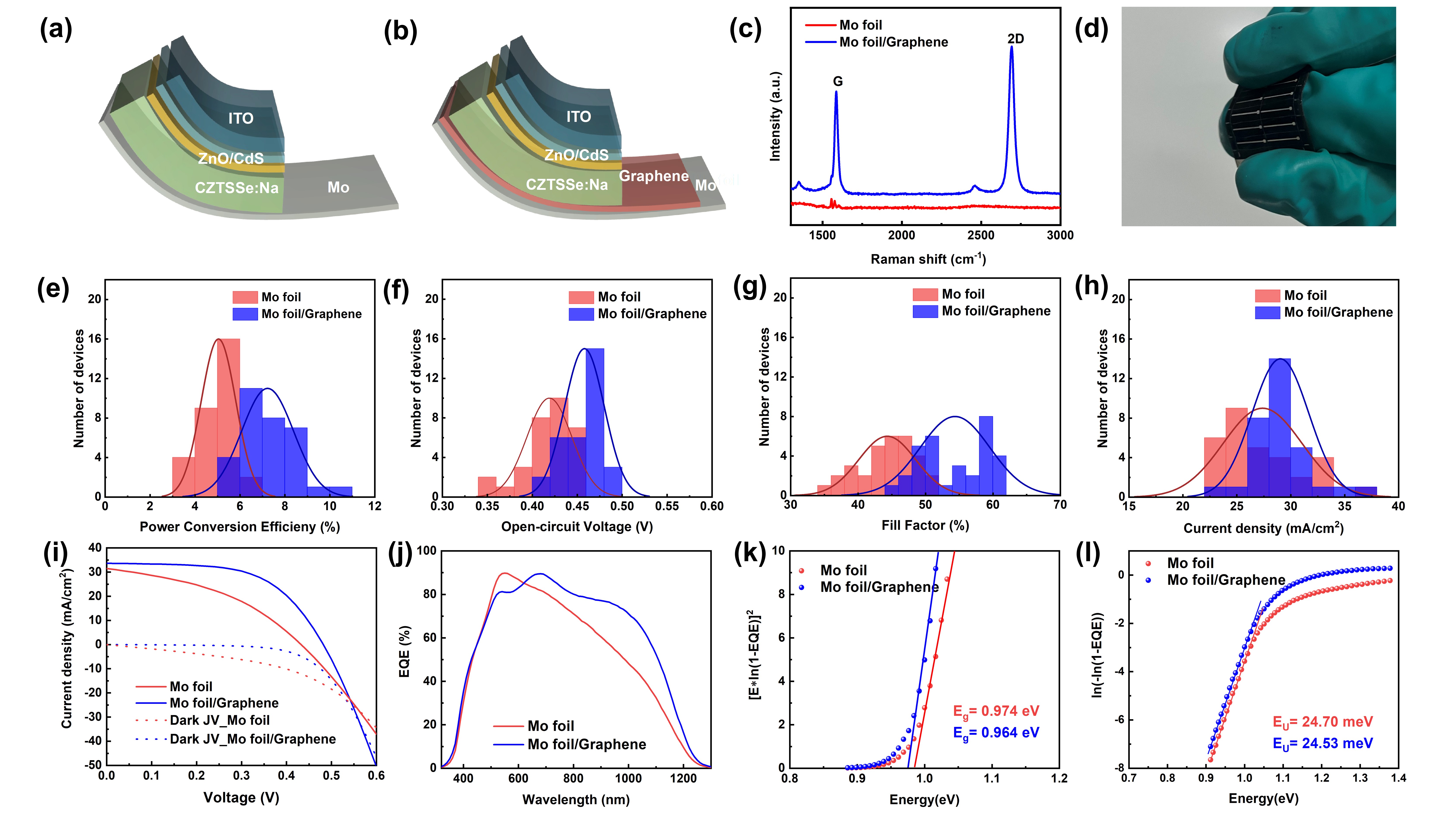}
    \caption[Performance of Flexible device structures without and with inserting graphene layers.]{Flexible device structures without and with inserting graphene layers: (a) Mo foil/CZTSSe/CdS/ZnO/ITO (denoted as 'Mo foil'), (b) Mo foil/Graphene/CZTSSe/CdS/ZnO/ITO (denoted as 'Mo foil/Graphene'). (c) The corresponding Raman spectra of Mo foil and Mo foil/Graphene substrates were collected with 532 nm excitation. Photovoltaic properties of Mo foil and Mo foil/Graphene devices: Histogram of (e) $V_{OC}$, (f)$J_{SC}$, (g)FF, and (h)PCE. (i) J-V curves of the champion device and corresponding (j) EQE, (k) calculated Eg, and (l) calculated Urbach energy.}
    \label{Figure 5-1}
\end{figure}

Configuration of a flexible CZTSSe solar cell with a structure of Mo foil/Graphene/CZTSSe/ CdS/ZnO/ITO is shown in Figure \ref{Figure 5-1}. The inserted graphene films were transferred onto Mo foil by a commonly used wet transfer method \cite{5-22}. Raman signals with a strong G-band peak at around 1580 cm$^{-1}$ and 2D-band peak at around  2700 cm$^{-1}$, with the intensity ratio of 2D and G band I$_{2D}$/I$_{D}$ clearly locates between 1 and 2 demonstrates the high quality of the transferred graphene films \cite{5-23}. Alkali metal (Na) was compensated by directly using a Na-contained CZTS precursor solution, which has already been shown to achieve high-performance CZTSSe solar cells on rigid substrates. The photovoltaic performance of flexible devices is shown in Figure \ref{Figure 5-1}(e-h), flexible CZTSSe devices without any inserting material as a comparison. It can be observed that the existence of graphene at the back interface helps to improve the PCE (Power conversion efficiency) by influencing V$_{OC}$ (open-circuit voltage), J$_{SC}$ (short-circuit current), and FF (fill factor). 

\begin{table}[ht]
\centering
\caption{The photovoltaic and diode parameters of the best Mo foil and Mo foil/Graphene solar cells. The values of E$_{g}$ are obtained from the EQE data. The diode parameters are extracted from dark J-V curves.}
\label{table 5-1}
\resizebox{0.9\columnwidth}{!}{%
\begin{tabular}{cccccc}
\hline
\textbf{Device} & \textbf{\begin{tabular}[c]{@{}c@{}}V$_{OC}$ \\ (V)\end{tabular}} & \textbf{\begin{tabular}[c]{@{}c@{}}J$_{SC}$\\ (mA/cm$^2$)\end{tabular}} & \textbf{\begin{tabular}[c]{@{}c@{}}FF \\ (\%)\end{tabular}} & \textbf{\begin{tabular}[c]{@{}c@{}}PCE \\ (\%)\end{tabular}} & \textbf{A} \\ \hline
Mo foil & 0.45 & 30.79 & 51.55 & 6.26 & 3.43 \\ \hline
Mo foil/G & 0.48 & 33.64 & 58.12 & 9.41 & 1.91 \\ \hline
\textbf{Device} & \textbf{\begin{tabular}[c]{@{}c@{}}J$_0$ \\ (mA/cm$^2$)\end{tabular}} & \textbf{\begin{tabular}[c]{@{}c@{}}R$_S$ \\ (ohm cm$^2$)\end{tabular}} & \textbf{\begin{tabular}[c]{@{}c@{}}G$_{SH}$\\ (mS/cm$^2$)\end{tabular}} & \textbf{\begin{tabular}[c]{@{}c@{}}Eg/q-V$_{OC}$\\ (eV)\end{tabular}} & \textbf{\begin{tabular}[c]{@{}c@{}}E$_U$\\ (meV)\end{tabular}} \\ \hline
Mo foil & $2.03 \times 10^{-4}$ & 2.45 & 17.40 & 0.524 & 24.70 \\ \hline
Mo foil/G & $1.72 \times 10^{-5}$ & 1.37 & 0.62 & 0.485 & 24.53 \\ \hline
\end{tabular}%
}
\end{table}

A comparison of J-V curves shows that the Mo foil/Graphene device has a higher JV crossover point, which implies increased series resistance and recombination in the Mo foil sample \cite{5-24}. EQE data shows a better response in longer wavelength regions of the Mo foil/Graphene sample. Both absorbers have similar band tailing properties from similar Urbach energy, identifying the enhancement of the Mo foil/Graphene sample originating from the change of back contact and the bottom region of the absorbers \cite{5-26}. 
Noticeably, the bandgap of the absorbers in flexible devices is smaller than that of rigid ones in our previous works, which implies an over-selenizing problem due to the prolonged Se diffusion in the selenization reaction. The thermal expansion of the Mo foil is physically larger than that in the Mo/glass samples, which provides more Se diffusion paths to the bottom during the process \cite{5-25}. Therefore, a thicker Mo(S, Se)$_{2}$ could be formed and severe back interface issues between Mo foil and CZTSSe absorber in flexible devices. 

\begin{figure}[H]
    \centering
    \includegraphics[width=1.\textwidth]{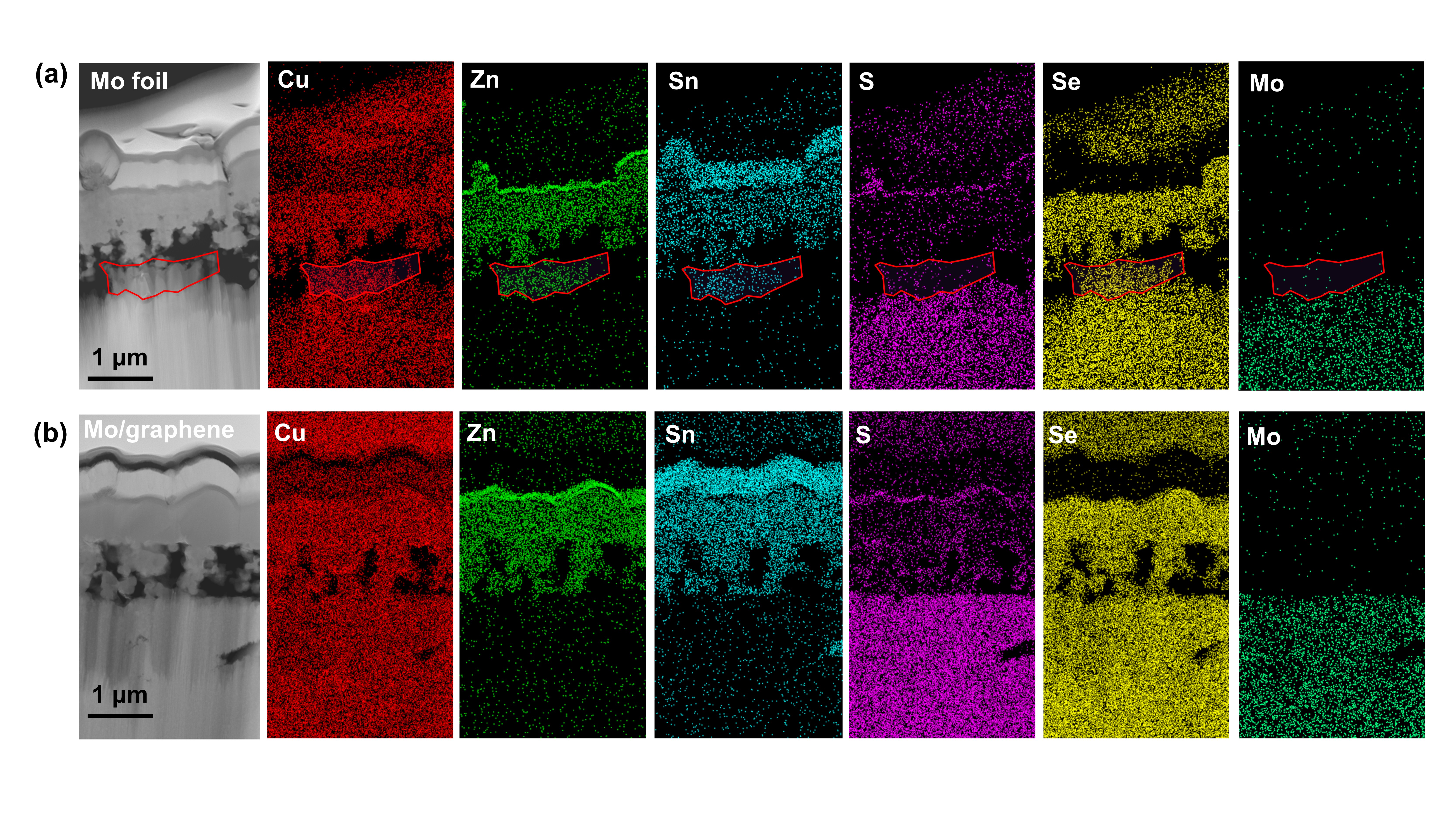}
    \caption[Cross-sectianal TEM and EDS mappings of the flexible CZTSSe solar cells with and without graphene inserted.]{Cross-sectianal TEM and EDS mappings of the elemental composition distribution of the flexible CZTSSe solar cells: (a) Mo foil device, (b) Mo foil/Graphene device. }
    \label{Figure 5-2}
\end{figure}

Distinctly different morphologies of the CZTSSe layers and the back interface can be observed in the two samples from cross-sectional TEM and EDS mapping images, shown in Figure \ref{Figure 5-2}. Mo foil device (in Figure \ref{Figure 5-2}(a)) has an uneven interface between Mo(S, Se)$_{2}$ and CZTSSe. There are many holes and substrate collapses in the bottom region, possibly caused by the excess expansion of Mo foil. Additionally, some of the CZTSSe grains (labeled by a red circle) have uneven attachment with Mo(S, Se)$_{2}$, due to the non-uniform chemical properties of the substrate surface which support the grain growth. However, different results were observed in the Mo foil/Graphene sample shown in Figure \ref{Figure 5-2}(b). The inserted graphene helps keep a smooth hetero-interface at the back contact, eliminating unwanted grain growth of CZTSSe started from the back interface. The surface of the Mo substrate was fully capped by two layers of graphene, and some dangling bonds and reactive chemical pots on the substrate were passivated by transferred graphene. Therefore, favorable grain growth of CZTSSe was realized, leading to better PV performance with enough light absorption and less recombination at the bottom region.

Calculations on thermal expansion have been done in the CZTSSe devices without and with graphene inserted during the selenization process. Different stages of the selenization reaction are simulated by using different Mo(S, Se)$_{2}$ thicknesses at the back contact. As shown in Figure \ref{Figure S5-1}, the existence of  Mo(S, Se)$_{2}$ at the beginning of the selenization reaction could induce the expansion of the Mo substrate, paving more Se diffusion paths, which might promote the formation of extremely thicker Mo(S, Se)$_{2}$ and cause the collapse of the back contact. Resulting of the calculation, the insertion of graphene could eliminate the thermal expansion caused by Mo(S, Se)$_{2}$, and keep a relatively thermal stable interface to support CZTSSe growth, even with the same thicknesses of Mo(S, Se)$_{2}$ at the interface. On the other hand, the thermal regulation effect from graphene will inhibit the formation of Mo(S, Se)$_{2}$ to some extent at the same time.

\begin{figure}[ht]
    \centering
    \includegraphics[width=\textwidth]{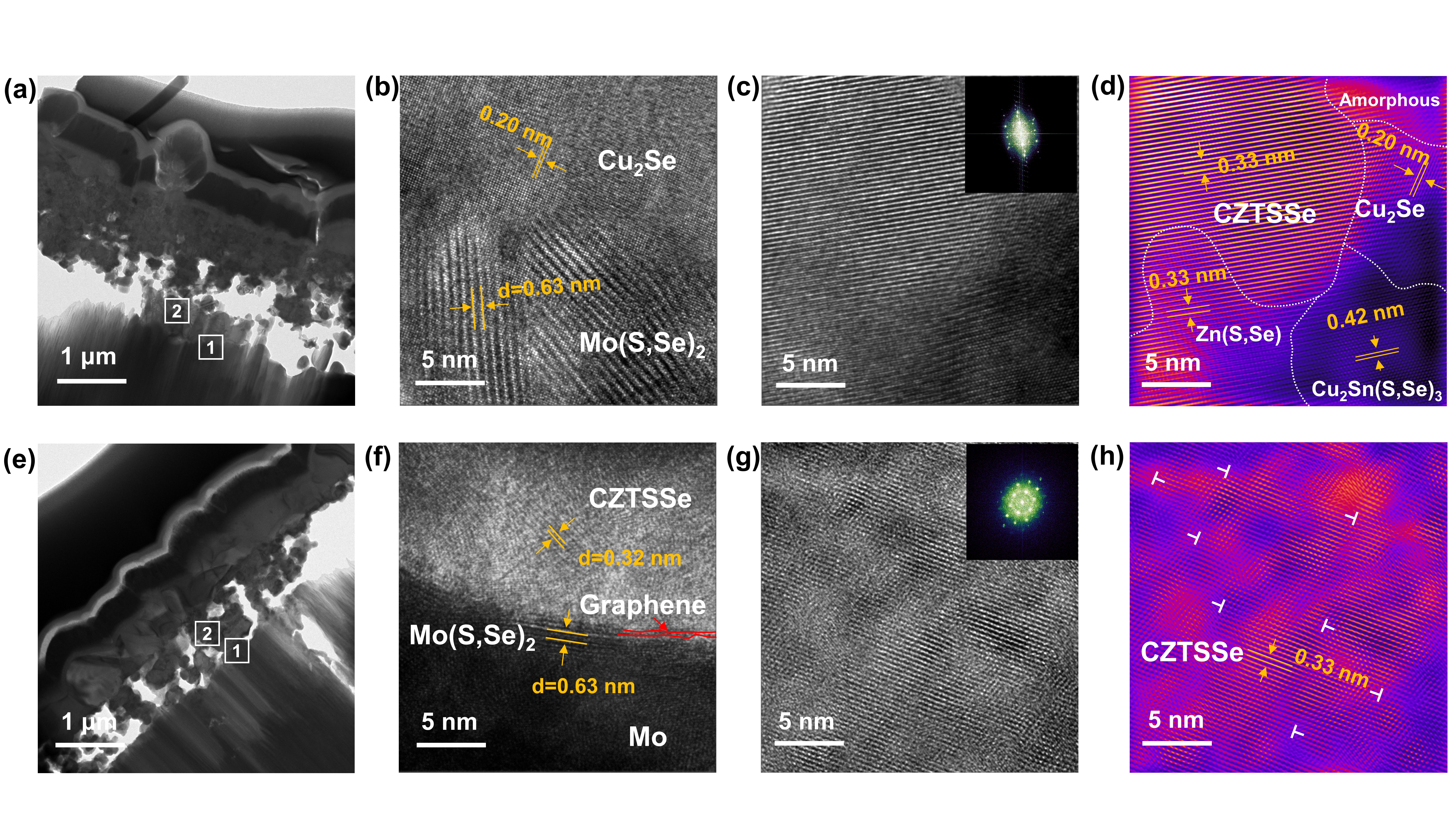}
    \caption[HRTEM images of two devices with and without graphene.]{Cross-sectional TEM image in bright field configuration of (a) Mo foil device and (e) Mo foil/Graphene device. HRTEM images of (b) region 1 and (c) region 2 in the Mo foil device, and (f) region 1 and (g) region 2 in the Mo foil/Graphene device. The insets in (c) and (g) are the corresponding FFT images. (d) and (h) are IFFI images of region 2 of two devices respectively}
    \label{Figure 5-3}
\end{figure}

\begin{figure}[H]
    \centering
    \includegraphics[width=\textwidth]{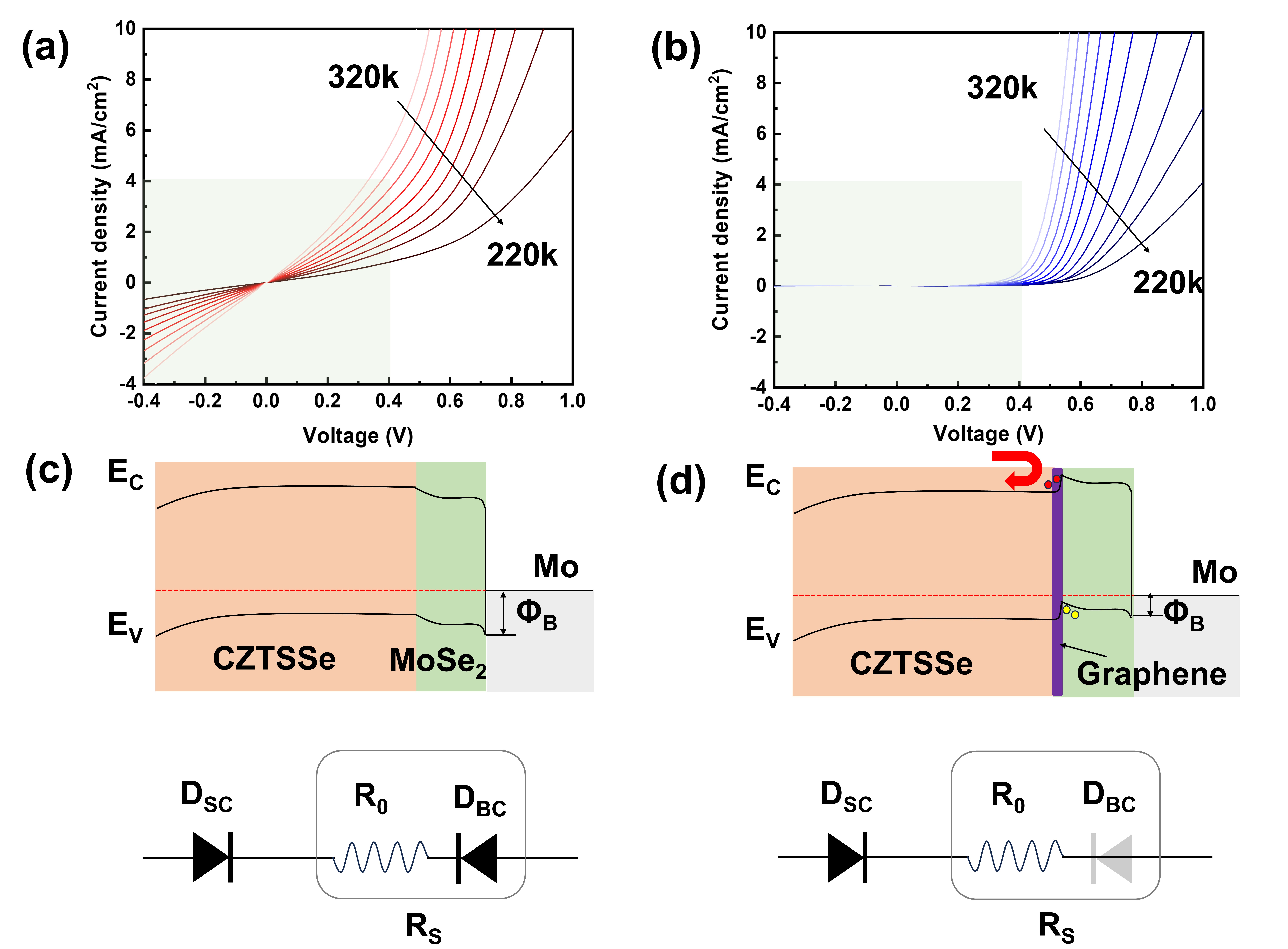}
    \caption[Temperature-dependent dark J-V curves and hypothetical circuit models.]{Temperature-dependent dark J-V curves of Mo foil (a) and Mo foil/Graphene (b) devices. Hypothetical back contact diagrams and circuit models are shown in (c) and (d). D$_{SC}$ is the solar cell diode. R$_{S}$ is series resistance, consisting of background or residual series resistance (R$_{0}$) and a blocking back contact diode D$_{BC}$.}
    \label{Figure 5-4}
\end{figure}

To investigate the atomistic effect of inserting graphene layers at the back contact interfaces of the two devices, high-resolution TEM (HRTEM) was performed. As shown in Figure \ref{Figure 5-3} (a) and (e), two regions were selected to give a more detailed discussion. The interfaces between CZTSSe and Mo(S, Se)$_{2}$, are labeled as "Region 1" in both of the samples. Vertical growth of Mo(S, Se)$_{2}$ can be seen in the Mo foil device with crystalline Cu$_{2}$Se deposited on top (shown in Figure \ref{Figure 5-4}(b)). The Cu$_{2}$Se possibly resulted from phase separation during the selenization reaction of the CZTSSe. In the Mo foil/Graphene sample, the bilayer graphene influenced the orientation of the Mo(S, Se)$_{2}$. The parallel growth of Mo(S, Se)$_{2}$ in (f) could influence carrier transport at the back contact \cite{5-27,5-28}. Additionally, the CZTSSe grown on the graphene indicated clear diffraction fringes, indicating better crystallinity and pure CZTSSe phase.

In "Region 2", slightly away from the back interface, two samples displayed distinct crytallinities. Mo foil samples exhibited distinguished secondary phases like Cu$_{2}$(S, Se), Zn(S, Se), and Cu$_{2}$Sn(S,Se)$_{3}$ clearly shown in IFFI image in Figure \ref{Figure 5-3}(d). In contrast, the Mo foil/Graphene sample showed a pure CZTSSe phase, with some tiny dislocations marked by white "T"s. Both the change in the orientation of Mo(S, Se)$_{2}$ and the crystallinity of the bottom regions of the absorber layer might significantly impact carrier extraction at the back interface of the solar cells.

\begin{equation}
\ dV/dJ=R_{\mathrm{S}}+nkT/q\left(J-G_{\mathrm{S}} V\right)
 \label{equation 5-1}
\end{equation}

Temperature-dependent measurements of dark J-V curves for two devices are provided in Figure \ref{Figure 5-4}. Series resistance of two devices are extracted and shown in Figure \ref{Figure S5-9}, by following the equation \ref{equation 5-1}, where $n$ is the diode ideality factor, $G_{S}$ is shunt resistance, $T$ is the temperature, $q$ is the electron charge, and $k$ is the Boltzmann constant respectively. The series resistance increases along with the temperature decrease from 320 to 240 K. The probable reason for series resistance increasing is the presence of a blocking contact barrier which is likely to occur at the CZTSSe and Mo interface. Note that graphene was inserted into the Mo foil/Graphene device, leading to a different electronic nature at the bottom Mo/CZTSSe contact. Such a difference could arise from the presence of a back-contact Schottky barrier in the Mo foil device that suppresses the hole transport across the Mo to absorber layer, while a quasi-ohmic contact in a Mo foil/graphene device. Band diagrams of the back contact are illustrated in Figure \ref{Figure 5-4}(c) and (d).

\begin{equation}
\ R_{\mathrm{BC}}=\frac{k}{q A^{*} T} \exp \left(\Phi_{\mathrm{B}} / k T\right)
 \label{equation 5-2}
\end{equation}

The barrier height of the back contact diode can also be derived from the temperature-dependent on dark J-V curves. The circuit models are also shown here, where consist of a main solar cell diode (D$_{SC}$), a back contact diode (D$_{BC}$), a background series resistance (R$_{0}$) due to the top layers and bulk resistance of CZTSSe absorbers. When the solar cell is forward biased, the back contact diode is in reverse bias and its conduction is limited by its reverse saturation current, which diminishes quickly at lower temperatures, thereby increasing the series resistance. Using this simple model, the barrier height of blocking contact can be estimated by fitting the equation \ref{equation 5-2}, where A$^*$ is the effective Richardson constant and $\Phi_{B}$ is the barrier height. The barrier height obtained from the Mo foil/Graphene device is 0.32 eV which is smaller than the 0.42 eV of the Mo foil device. The presence of the low barrier value is considered to be closer to an ohmic contact, consistent with its low R$_s$ and high FF.

\begin{figure}[ht]
    \centering
    \includegraphics[width=1.\textwidth]{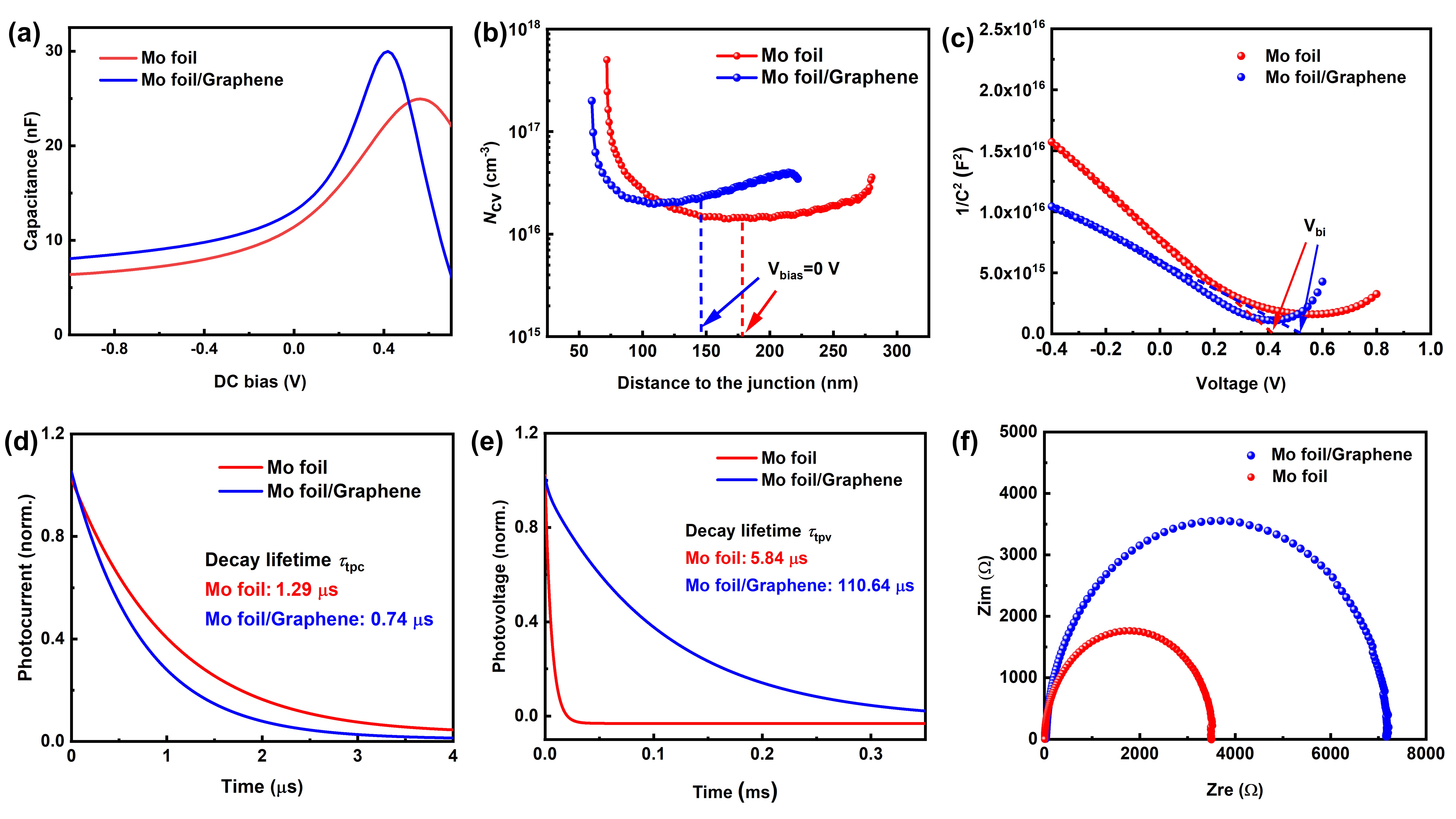}
    \caption[Charge transfer and recombinations.]{(a) C-V curves. (b) CV curves and the calculated width of the space charge region are shown as the dashed lines at 0 V bias. (c) Plot of 1/C$2$ versus voltage showing the variation in the built-in potential, V$_{bi}$. Normalized TPC spectra in (d) and normalized TPV spectra in (e). Impedance spectra for the Mo foil and Mo foil/Graphene devices.}
    \label{Figure 5-5}
\end{figure}

The carrier collection within two devices was investigated by conducting Capacity-voltage (C-V) and charge dynamics analysis to elucidate the underlying device physics. Capacity-voltage (C-V) and derived depletion region width are shown in Figure \ref{Figure 5-5}. With enhanced crystallinity and phase purity at the bottom region of the absorber, the carrier density derived from C-V has been improved from $1.44 \times 10^{16}$ cm$^{-3}$ to $2.26 \times 10^{16}$ cm$^{-3}$, which might contribute to enhanced conductivity. The depletion region was changed from 178.9 nm to 145.66 nm, suggesting reduced recombination losses. A larger built-in potential of around 0.51 V is the reason for higher open-circuit voltage and overall efficiency \cite{5-29}.

Transient photocurrent (TPC) and transient photovoltage (TPV) measurements are used to investigate the charge transfer and recombination characteristics of the two devices. TPC measures the decay of photocurrent in a closed circuit, reflecting the transport properties of photogenerated carriers in the device, while TPV measures the decay of photovoltage in an open circuit, indicating recombination properties. From TPV decay, the charge recombination lifetime ($\tau_{tpv}$) is determined. The $\tau_{tpv}$ values for Mo foil and Mo foil/Graphene devices are 5.84 and 110.64 $\upmu$s, respectively, showing better recombination suppression in the Mo foil/Graphene device. Similarly, TPC decay gives the charge transfer lifetime ($\tau_{tpc}$), indicating the time for carriers to reach the electrode. The $\tau_{tpc}$ values for Mo foil and Mo foil/Graphene devices are 1.29 and 0.74 $\upmu$s, respectively, implying improved charge transport capability in the Mo foil/Graphene device. Therefore, the insertion of graphene enhances both carrier recombination suppression and transport in CZTSSe films.


\begin{table}[]
\caption{Summary of the results derived from C-V, TPV, TPC, and EIS measurements.}
  \label{Table_2}
\begin{tabular}{ccccccc}
\hline
\textbf{Device} & \textbf{\begin{tabular}[c]{@{}c@{}}V$_{bi}$\\ (V)\end{tabular}} & \textbf{\begin{tabular}[c]{@{}c@{}}N$_{CV}$\\ (cm$^{-3}$)\end{tabular}} & \textbf{\begin{tabular}[c]{@{}c@{}}Depletion width\\ (nm)\end{tabular}} &
\textbf{\begin{tabular}[c]{@{}c@{}}$\tau_{tpv}$\\ ($\mu$s)\end{tabular}} & \textbf{\begin{tabular}[c]{@{}c@{}}$\tau_{tpc}$\\ ($\mu$s)\end{tabular}} & \textbf{\begin{tabular}[c]{@{}c@{}}$R_{rec}$\\ (k$\Omega$)\end{tabular}} 
\\ \hline
Mo foil & 0.41 & $1.44 \times 10^{16}$ & 178.90 & 5.84 & 1.29 & 3.53 \\ \hline
Mo foil/G & 0.51 & $2.26 \times 10^{16}$ & 145.66 & 110.64 & 0.74 & 7.17 \\ \hline
\end{tabular}
\end{table}

Moreover, electrochemical impedance spectroscopy (EIS) was conducted to analyze the carrier recombination behavior. The Nyquist diagrams of EIS spectra and equivalent circuit diagrams of the devices are described in Figure \ref{Figure 5-5}(f). The intersection points of high and low frequency with the x-axis are the series resistance (R$_{s}$) and recombination resistance (R$_{rec}$), respectively, while the shunt resistance (R$_{sh}$) is represented by the diameter of the impedance spectrum circle line. The R$_{rec}$ values of the Mo foil and Mo foil/Graphene devices are 3.53 and 7.1 K$\Omega$ respectively. The smaller R$_{rec}$ of the Mo foil/Graphene device is consistent with the TPV result, leading to effective suppression of defect recombination at the interface.

\subsection{Conclusion}
In this work, we investigated the impact of inserting graphene layers at the Mo foil/CZTSSe interface and studied their influences on CZTSSe grain growth and the performance of flexible CZTSSe solar cells. Samples with a graphene interlayer show a smoother and denser morphology, with significantly improved crystallinity of CZTSSe. The champion flexible CZTSSe thin-film device, incorporating a graphene back-interface passivation layer, achieved a conversion efficiency of 9.41\%, an open-circuit voltage of 0.48V, a short-circuit current density of 33.64 mA/cm$^2$ and a fill factor of 58.12\%.

Detailed analysis revealed that the insertion of the graphene not only influenced the grain growth on the Mo foil substrate but also prohibited the phase decomposition at the nanoscale of the back contact and the bottom region of the absorbers. Defects within absorbers were significantly passivated due to enhanced crystallinity. Carrier collection at the back contact was more effective with better physical contact built by the inserted graphene. A lower potential barrier was obtained, facilitating hole transport and mitigating the reverse diode effect at the back interface, leading to the formation of quasi-ohmic contact at the back contact.


\newpage
\bibliography{achemso-demo}

\end{document}